\documentclass[9pt]{article}
\pdfoutput=1
\usepackage{spconf,amsmath,graphicx, array, subfig, setspace, textcomp, verbatim, amsmath}
\usepackage{scrextend, bm}
\newcolumntype{M}[1]{>{\centering\arraybackslash}m{#1}}

\title{Cross-Modality Distillation: A case for Conditional Generative Adversarial Networks}
\name{Siddharth Roheda$ ^{\star} $ \qquad Benjamin S. Riggan$ ^{\dagger} $ \qquad Hamid Krim$ ^{\star} $ \qquad Liyi Dai$^{\ddagger}$} 

\address{ $ ^{\star} $Department of Electrical and Computer Engineering, North Carolina State University, Raleigh, NC, USA\\ $ ^{\dagger} $U.S. Army Research Laboratory, Adelphi, MD, USA\\ $ ^{\ddagger}$U.S. Army Research Office, 800 Park Offices Dr., Durham, NC, USA }
\begin{document}
%
\maketitle
\begingroup\makeatletter\def\f@size{9}\check@mathfonts
\setlength{\abovedisplayskip}{1.5pt}
\setlength{\belowdisplayskip}{1pt}
\begin{abstract}
In this paper, we propose to use a Conditional Generative Adversarial Network (CGAN) for distilling (i.e. transferring) knowledge from sensor data and enhancing low-resolution target detection. In unconstrained surveillance settings, sensor measurements are often noisy, degraded, corrupted, and even missing/absent, thereby presenting a significant problem for multi-modal fusion. We therefore specifically tackle the problem of a missing modality in our attempt to propose an algorithm based on CGANs to generate representative information from the missing modalities when given some other available modalities. Despite modality gaps, we show that one can distill knowledge from one set of modalities to another. Moreover, we demonstrate that it achieves better performance than traditional approaches and recent teacher-student models.  
\end{abstract}
\begin{keywords}
Missing Modalities, Generative Adversarial Networks, Target Detection, Multi-Modal Fusion
\end{keywords}
\section{Introduction}
\label{sec:intro}

Some sensor measurements are often noisy, missing, or unusable in unconstrained surveillance settings, and sometimes there are limitations on the types of sensors that are deployed.
In this scenario, it is common to ignore such sensors, with a potential negative impact on performance relative to the ideal scenario (i.e. all sensors are available and functional). We consider exploiting prior knowledge about the relationship between the two sets of modalities, so that a detection system can safeguard a high detection accuracy.

There has recently been an interest in this type of knowledge
transfer. Shao et al. \cite{shao2015sparse} proposes to use multiple deep auto-encoders with a bagging strategy and Robust PCA to address missing modalities for image classification. Hoffman et al. \cite{hoffman2016learning} introduces hallucination networks that can address a missing modality at test time by distilling knowledge into the network during the training. Furthermore, \cite{rastegar2016mdl} shows that considering interactions between modalities may lead to a better representation. Recent work in Domain Adaptation \cite{gopalan2011domain, zheng2012grassmann, ni2013subspace} addresses differences between source and target domains. In these works, the authors try to learn a set of intermediate domains (represented as points on the Grassman manifold in \cite{gopalan2011domain, zheng2012grassmann} and by dictionaries in \cite{ni2013subspace}) between the source domain and the target domain. These approaches to cross-modal inference do not explore the distillation of information between modalities with different phenomenologies (e.g., imaging and acoustic data). In this paper, we try to explore the possibility of such a transfer of knowledge by leveraging information from seismic and acoustic sensors while training a model that takes in images for classification.


\textbf{Our Contributions:} We propose to use Conditional Generative Adversarial Networks (CGANs) \cite{mirza2014conditional} to generate representative information from the missing sensor modalities (e.g. temporal data) while conditioned on the available modalities (e.g. spatial data). This, in effect, distills knowledge from the missing modalities into the model trained for dealing with available modalities.
We modify the generator cost function so that the model learns to extract realistic and discriminative information corresponding to the missing modalities. Furthermore, we compare our approach to traditional approaches (i.e. ignore missing modalities) and recent distillation approaches. We also evaluate various cost functionals in order to better understand the impact of different loss functions and regularizations.
\section{Background and Related work}


\subsection{Teacher-Student Distillation Model}

This model has been largely used for compressing large neural networks (i.e. teacher) into smaller ones (i.e. students), while achieving a similar performance. In \cite{buciluǎ2006model}, the output from the teacher was used as the target probability distribution for the student. Hence, the student was able to achieve similar performance to the teacher. Further improvements on the technique introduced in \cite{buciluǎ2006model} are proposed in \cite{hinton2015distilling}. In \cite{vapnik2015learning}, a type of teacher-student model plus additional privileged information is used. This kind of distillation was then extended to training networks to deal with missing modalities in \cite{hoffman2016learning}, where L\textsuperscript{2} loss (hallucination loss) is used to train a hallucination network.

\subsection{Generative Adversarial Networks}

Generative Adversarial Networks (GANs) were recently introduced in \cite{goodfellow2014generative}. GANs are composed of two parts: a generator and discriminator.  The generator aims to confuse the discriminator by randomly synthesizing a realistic sample from some underlying distribution. The adversarial discriminative network aims to differentiate between actual samples and generated ones. When the generator and discriminator networks were alternatively trained, the generator was able to successfully generate a random, but realistic sample from the distribution of the training data.  Later, Conditional Generative Adversarial Networks (CGANs) in \cite{mirza2014conditional} were trained to generate samples from a class conditional distribution. Here, the input to the generator was replaced by some useful information rather than random noise. Therefore, the objective of the generator was to generate realistic data, given the conditional information.
CGANs have been used to generate random faces given attributes \cite{gauthier2014conditional} and to produce relevant images given a text descriptions \cite{reed2016generative}. 

\subsection{Noise Models} \label{noisemodels}
Past works have explored various noise models for CGANs. Without a random process (i.e. noise), the generator would produce deterministic results. In \cite{wang2016generative} a Gaussian input noise is provided to the generator. But often, the generator can learn to ignore this noise \cite{mathieu2015deep, isola2016image}. Another noise model that has been used is dropout noise in \cite{isola2016image}, where dropout is applied to several layers of the generator. 

\section{Problem Formulation}
Consider $m$ sensors surveilling an area of interest. We wish to detect the presence of a target, given the data collected by these sensors. Let the observations for the $ i^{th} $ sensor be denoted by, $ \bm{a}_i = \{a_{ij}\}_{j = 1, ..., n} $, where, $ a_{ij} $ corresponds to the signal value at time $j$. Now, assume that observations from $ k $ sensors, $ \bm{A} = \{\bm{a}_1, \bm{a}_2, ..., \bm{a}_k\} $, are missing or unusable during the testing phase, while the rest of the sensor observations, $ \bm{B} = \{\bm{a}_{k+1}, \bm{a}_{k+2}, ..., \bm{a}_m\} $, are usable for target detection. Let $ M: \bm{A} \to F_M $ be a mapping from the data (available for training), $ \bm{A} $, to a feature space, $ F_M $. Let $C_{\bm{A}}(F_M) = \bm{w}_{\bm{A}}^T F_M + b_{\bm{A}}$ be a scoring function gives a detection score, where, $\bm{w}_{\bm{A}} $ and $ b_{\bm{A}} $ is the weight vector and the bias for a classifier trained to detect a person. Then, the probability of detection is $ P_{\bm{A}} = S(C_{\bm{A}}(F_M)) $, where, $S(x) = 1/(1-e^{-x})$, is the sigmoid function. Also consider another mapping, $ G: \{z, \bm{B}\} \to F_G $, that maps random noise, z, and the usable data, $ \bm{B} $, to a feature space, $ F_G $. 

Our goal is to then use the available sensor data, $\bm{B}$, in order to predict features that would have been generated by $\bm{A}$, and hence improve detection performance with the usable data. That is, we would like to find the mapping, $ G $, such that $ F_G = G(z, \bm{B}) \approx F_M = M(\bm{A}) $.


\section{Proposed Method}

Fig. \ref{block} shows a high level block diagram for learning the mapping $G$.
The left side of the diagram depicts how $\bm{A}$ is mapped to the feature space, $F_M $, by the mapping $M$. These features are then used for classification purposes, and a linear classifier, with scoring function, $C_{\bm{A}}(.)$, is trained using $F_M $. 

\begin{figure}[!h]
	\centering
	\includegraphics[width=0.37\textwidth, height=0.15\textheight]{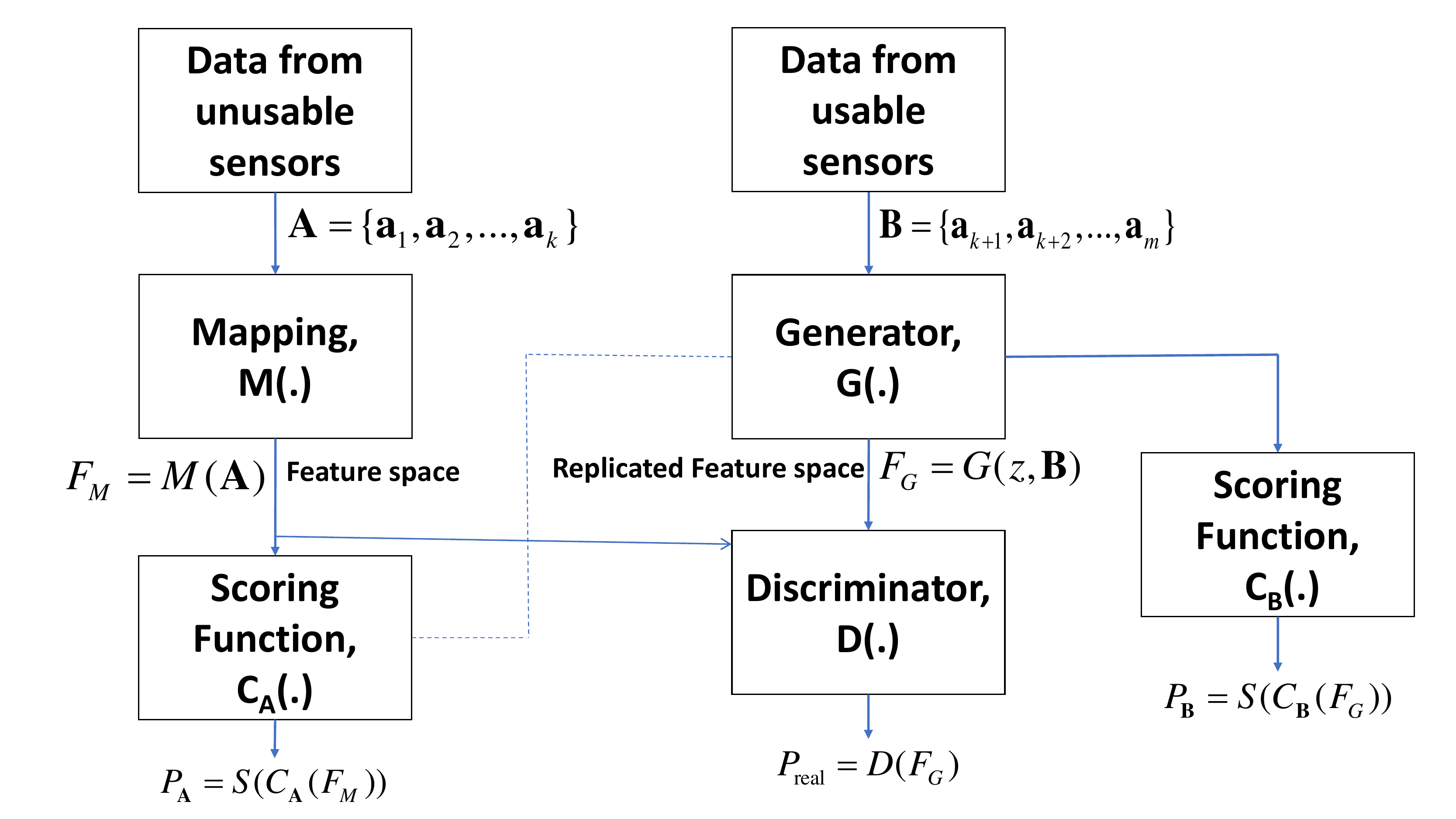}
	\caption{\small A high level block diagram for training the generator}
	\label{block}
\end{figure}


On the right side of the diagram, we have a CGAN network that we use to replicate the feature space, $ F_M $. The generator from the CGAN network is used as the mapping, $G$, and takes the data from available modalities as the input. This generative model is pitted against an adversary: a discriminator function, $D(.)$, that tries to discriminate between the generated feature space, $F_G$, and the targeted feature space, $F_M$. The generator tries to produce a feature space, $ F_G $, that replicates $F_M$ as closely as possible and causes misclassification by the discriminator. Then using the CGAN formulation as defined in \cite{mirza2014conditional, isola2016image}, we have,
\begin{equation} 
\min_{G(.)} \max_{D(.)} V(D,G),
\end{equation}
where
\begingroup\makeatletter\def\f@size{9}\check@mathfonts
\begin{equation}
	\begin{split}	
V(D,G) = {\rm I\!E}_{\bm{B} \text{\texttildelow} p_{data}(\bm{B}), M(\bm{A}) \text{\texttildelow} p_{data}(M(\bm{A}))} [log D(\bm{B}, M(\bm{A}))] + \\ {\rm I\!E}_{{\bm{B} \text{\texttildelow} p_{data}(\bm{B})}, {z \text{\texttildelow} p_z(z)}} [log (1 - D(\bm{B}, G(z, \bm{B})))]
\label{CGANeq},
	\end{split}
\end{equation}
\endgroup
and $ {\rm I\!E}(.) $ is the expectation operator with respect to the observed data. This, in effect, minimizes some distance loss function, $ \mathcal{L}(F_G, F_M) $, such that, $F_G$ and $ F_M $ are close to each other, leading to performance comparable to that achievable prior to missing $\bm{A}$. While it is reasonable to expect good classification performance from such an attempt, we found that the discriminative information is suboptimal. This results in the discriminator adapting to ``erroneous real features" and lead to an undesirable outcome. This may be attributed to a convergence to the wrong stable point of the objective functional. As an illustration, when considering an image patch in which very few pixels correspond to the target (see Fig. \ref{imagesamples}), the generator may be easily confused. 
Therefore, we require additional criteria to guide the generator to a better convergence point.

We remedy this issue by letting the classifier (with the scoring function $ C_{\bm{A}}(.) $)  influence $G(.)$. This classifier is useful for guiding the generator to produce more discriminative features derived from $ \bm{B} $. This causes the generator to produce features that lie on the correct side of the decision boundary. Adding this classifier influence does not have a large impact on training time as the classifier is pre-trained. The cross-entropy loss for a  classifier with a scoring function, $C(.)$, given a set of features, $ X = \{\bm{x}_i\}$, with classification labels, $ L = \{l_i\} $, where index $ i $ represents the $ i^{th} $ sample, is given as,
\begin{equation}
	C_{Loss}(X,L) = \sum_i - l_i log S(C(\bm{x}_i)).
\end{equation}
We further incorporate the minimization of some measure of distance between the generated features and the real features. For our implementation, we minimize the $L^2$ loss: $L^2(F_M,F_G)=\sum_i (f_{M_i} - f_{G_i})^2$, where $f_{M_i}$ and $f_{G_i}$ denote samples from $F_M$ and $F_G$, respectively. This loss encourages the generator to produce a representation corresponding to the conditional input.
Since we used the classifier with scoring function $ C_{\bm{A}}(.) $ to influence the training of the generator, $ G(.) $ , it is reasonable to expect its good performance in classifying features generated by $G(.)$. $F_G$ is unfortunately not an exact copy of $F_M$, albeit close. So, a better classifier for $ F_G $ may exist, and is shown as the scoring function $C_{\bm{B}}(.)$ in the block diagram in Fig. \ref{block}.  We use dropout noise as the input noise to the generator, which is discussed in \cite{isola2016image}. Let $U$ be the set of class labels for the training samples, $C_{{\bm{A}}_{Loss}}(F_G, U)$ be the classification loss for the generated features using the pre-trained classifier with scoring function, $C_{\bm{A}}(.)$, and $L^2(F_M,F_G)$ be the $L^2$ loss between the targeted features and the generated features. Then, we formulate the following optimization problem to efficiently replicate $ F_M $,
\begin{equation}
	\min_{G(.)}\max_{D(.)} J(D,G) = V(D,G) + \alpha C_{\bm{A}_{Loss}}(F_G, U) + \beta L^2(F_M, F_G)
\end{equation}
Where, $ \alpha \text{ and } \beta $ are scaling coefficients for controlling influence of the pre-trained classifier and the $L^2$ loss. For our implementation, we use $\alpha = 10^{-3}$, and $ \beta = 10^{-5} $.

\section{Experiments and Results}


\subsection{Experimental Setup}

For our experiments we used pre-collected data from a network of seismic sensors, and acoustic sensors deployed in a field, where people were walking around in specified patterns. Details about this sensor setup and experiments can be found in \cite{nabritt2015personnel}. This dataset has been previously used for target detection in \cite{lee2017accumulative, leeoptimized}, where the authors focus on detection/classification using seismic and acoustic sensors only. Some data samples from the dataset can be seen in Fig. \ref{imagesamples}. It can be seen that while the response from seismic and acoustic sensors has significant discriminative information, the same cannot be said for the images from the video camera. So, in an ideal case, we would use the responses from the seismic and acoustic sensors to detect target presence. We assume those sensors are unavailable during the testing phase. 
\begin{figure}[!h] 
	\centering
	\includegraphics[width = 0.29\textwidth]{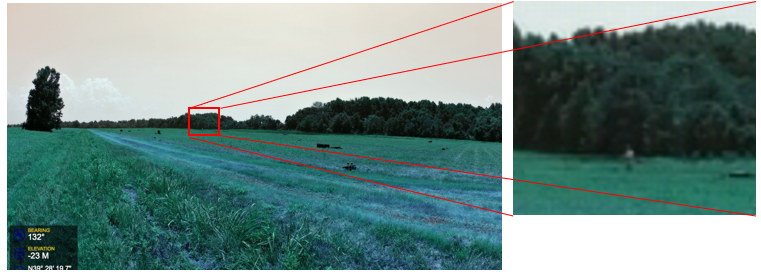} \hfill
	\includegraphics[width = 0.11\textwidth,height=0.08\textheight]{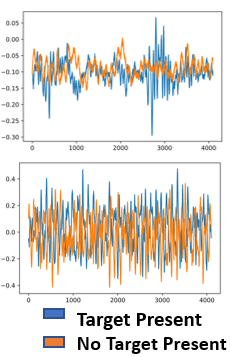} 
	\caption{\small A sample video frame with a small human target (left),  and samples from seismic (top right) and acoustic (bottom right) sensors}
	\label{imagesamples}
\end{figure}
	\begin{table*} [tbp]
	\begin{center}	
		\begin{tabular}{|M{8.1cm}| M{1.87cm}|M{1.87cm}| M{1.85cm}| M{1.85cm}|}
			\hline
			\textbf{Method} & \textbf{\small Avg Accuracy (using $C_{\bm{A}}(.)$)} & \textbf{\small Avg Accuracy (using $C_{\bm{B}}(.)$)} & \textbf{\small Avg F-score (using $C_{\bm{A}}(.)$)} &\textbf{\small Avg F-Score (using $C_{\bm{B}}(.)$)}\\
			\hline
			\small Seismic, Acoustic Sensors\footnotemark{} & \small 95.93\% & - & \small 0.91 & - \\
			\hline
			\small Video Frames & - & \small 76.67\% & - & \small 0.53\\
			\hline
			\small Hallucination Networks (Teacher-Student Model) & - & \small 82.40\% & -  & \small 0.63\\
			\hline
			\small Cross-Modal Distillation (CGAN) & \small 67.10 \% & \small 67.73\% & \small 0.59 &  \small 0.60\\
			\hline
			\small Cross-Modal Distillation (CGAN + Classifier Loss) & \small 90.93\% & \small 91.33\% & \small 0.81 & 0.83\\
			\hline
			\small \textbf{Cross-Modal Distillation (CGAN+ Classifier Loss + $\mathbf{L^2}$)} & \small 92.18\% & \small \textbf{93.52}\% & \small 0.87 & \small \textbf{0.88}\\
			\hline
			\small Cross-Modal Distillation (CGAN + $L^2$) & \small 89.21\% & \small 90.89\% & \small 0.73 & \small 0.76\\
			\hline
			\small Cross-Modal Distillation (Classifier Loss + $L^2$) & \small 85.97\% & \small 86.01\% & \small 0.70  & \small 0.72\\
			\hline
			\small Cross-Modal Distillation (Distance Correlation \cite{szekely2009brownian} + Classifier Loss) & \small 87.19\% & \small 91.11\% & \small 0.77 & \small 0.81\\
			\hline
			\small Cross-Modal Distillation (CGAN + Distance Correlation \cite{szekely2009brownian} + Classifier Loss) & \small 90.21\% & \small 90.9\% & \small 0.80 & \small 0.80\\
			\hline
		\end{tabular}
	\end{center}
	\caption[Caption for LOF]{\small Performance Comparison \footnotemark{} }
	\label{acc_table}
	
\end{table*} 

\subsection{Implementation Details} \label{implementation}

To simplify the generator conditioned on images, we feed image patches of size 64x64 instead of the entire 720x1080 image. This makes the conditioning on patches of the image, and the generator generates features that are similar to those from the network trained over the seismic and acoustic data. An image patch including a person is labeled as a target and correlated with 1 second long (4096 samples) seismic and acoustic data from the closest node to the person. On the other hand, the person free patches, are labeled as non-target and correlated with 1 second long (4096 samples) seismic and acoustic data from the farthest node to the person. 

The mappings, $ M(.) \text{ and } G(.) $ are approximated by neural networks. The network structure used is the same across all algorithms that are compared. The structures are identical for the mapping of seismic, acoustic data into the feature space and for generation of features from image patches, except for the fact that the former uses 1-d convolutional filters, while the latter uses 2-d convolutional filters. A 3 layered CNN structure is used, with 2x2 max-pooling after the first and last convolutional layers. The first layer uses a filter size of 3 and the last two layers approximate a layer of filter size 5 by using two layers of filter size 3. As discussed in \cite{szegedy2016rethinking}, using two layers of filter size 3 is computationally faster, with an equivalent output to using a single layer with filter size 5. The convolutional layers are followed by a fully connected layer that transforms the output of the convolutional layers into a feature vector. The fully connected layer generates a feature vector with 50 values, before being fed into a classification layer. This is the feature vector we try to predict using the CGAN based algorithm we propose.  
\subsection{Results}

The results for the proposed algorithm can be seen in Fig. \ref{results}, in comparison to a network that was trained over just video frames. It can be seen that the network trained over video frames only biases toward the horizon in the image, where the person is mostly found to be walking, but this issue is resolved when cross-modal distillation is performed, and a more focused detection is achieved. Fig. \ref{results_diff_angle} shows results on a video frame captured from a different angle, and shows that the proposed algorithm is not over trained, and is robust to different perspective.

\begin{figure}[tbp] 
	\includegraphics[width = 0.14\textwidth]{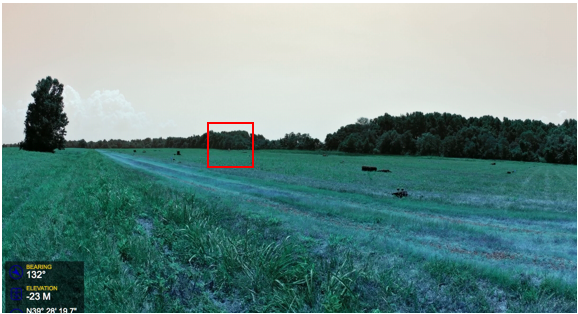} \hfill
	\includegraphics[width = 0.14\textwidth]{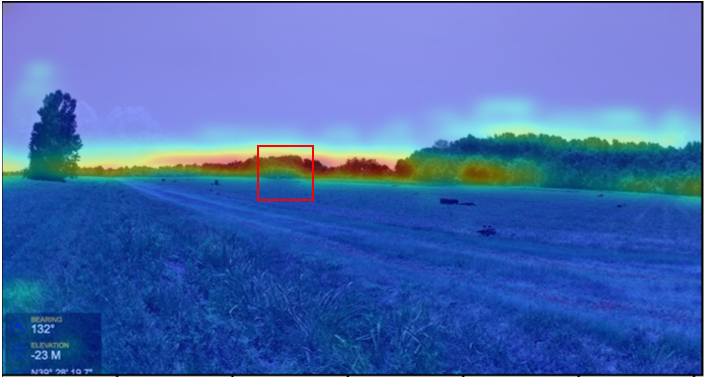} \hfill
	\includegraphics[width = 0.14\textwidth]{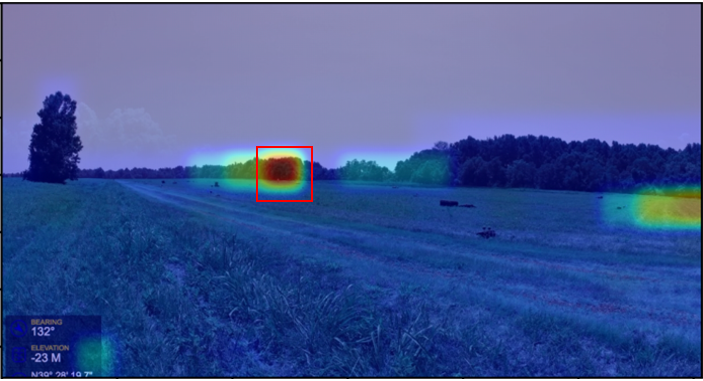} 
	\includegraphics[height= 0.075\textwidth]{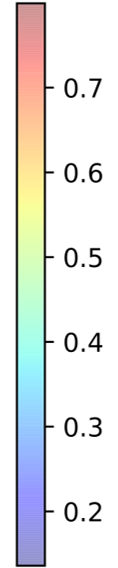} \hfill	
	\caption{\small Original video frame (left), Detection Probability when network was trained using just the video frames (center), Detection Probability when network was trained using proposed Cross-Modal distillation (right). Red box marks true location of target}
	\label{results}
\end{figure}
\begin{figure}[tbp] 
	\centering
	\includegraphics[width = 0.2\textwidth, height=0.06\textheight]{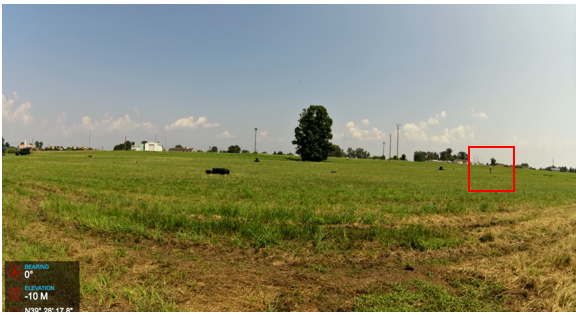} \hfill
	\includegraphics[width = 0.2\textwidth, height=0.06\textheight]{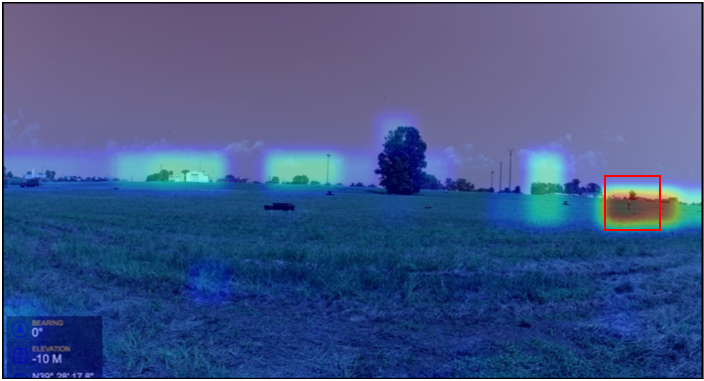} 
	\includegraphics[height= 0.08\textwidth]{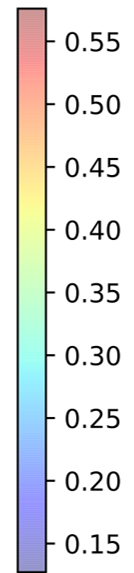} \hfill	
	\caption{\small Detection results on a video frame captured from a different angle shows robustness of the proposed algorithm}
	\label{results_diff_angle}
\end{figure}

Further, Fig. \ref{conv_layer_results} shows the output from the convolutional layer when a patch from the input video frame was fed in. It can be seen that the network trained using just the video frames is highly responsive along the edge between trees and grass. While, this is a feature, it is not of interest here. On the other hand, the network trained using cross-modal distillation is highly responsive around the head of the person in the image, which is a much more interesting feature for our purpose.
\begin{figure}[tbp] 
	\includegraphics[width = 0.102\textwidth, height=0.05\textheight]{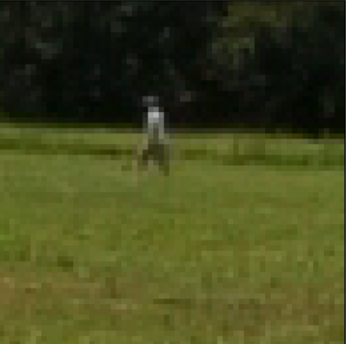} \hfill
	\includegraphics[width = 0.10\textwidth, height=0.05\textheight]{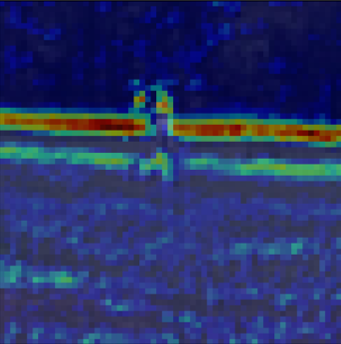} 
	\includegraphics[height= 0.065\textwidth]{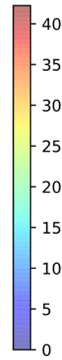} \hfill
	\includegraphics[width = 0.10\textwidth, height=0.05\textheight]{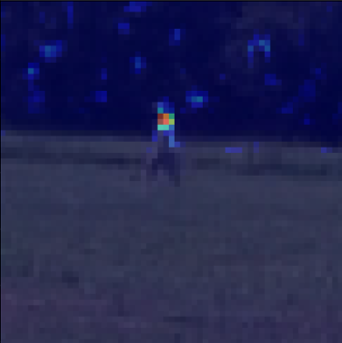} 
	\includegraphics[height= 0.065\textwidth]{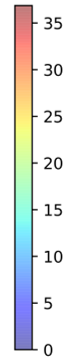} \hfill	
	\caption{\small Positive Patch (left), Output from convolutional layer trained using video frames only (center), Output from convolutional layer trained using proposed Cross-Modal distillation (right)}
	\label{conv_layer_results}
\end{figure}
Table \ref{acc_table} presents the detection accuracies for different combinations of cost functions. Accuracies and F-scores are reported for both cases of classifiers, i.e. the pre-trained classifier, $   C_{\bm{A}}(.) $, and the finetuned (on the generated feature space) classifier, $ C_{\bm{B}}(.) $. Detection performance achieved by the seismic and acoustic sensor data is optimal. But, when these sensors are damaged or not available, we must use the available video frames for target detection. On comparing the performance of classifier based on only video frames with other distillation techniques, it is clear that some transfer of knowledge definitely takes place, and helps us build a model with improved detection performance. Furthermore, our approach, which is based on using CGANs along with additional losses, is able to improve the knowledge transfer compared to teacher-student model based distillation. From Table \ref{acc_table} we can see that the combination of CGAN conditioned on video frames, a classifier influence, and an $L^2$ loss yields the best detection performance, which is very close to that achieved by the seismic and acoustic sensors.
\addtocounter{footnote}{-2}
\stepcounter{footnote}\footnotetext{This is the ideal performance, which is not achievable since we assume these sensors are missing.}
\stepcounter{footnote}\footnotetext{Cells with a ``-'' cannot be reported due to incompatibilities. }

\section{Conclusion}
In this paper we proposed a technique for cross-modal distillation that uses Conditional Generative Adversarial Networks (CGANs) in order to predict features from modalities that may be unusable for testing/implementation due to damage or other potential limitations. This cross-modal distillation allows us to leverage information from these missing modalities in order to guide the model being trained to learn highly discriminative features. Experiments show that detection using modalities with low discriminative information (due to low resolution of target in our case) can be significantly improved by distilling knowledge from a modality with higher discriminative power. This algorithm improves upon the performance using recent Teacher-Student model for distillation, and hence provides robustness in presence of limited sensing modalities.

\endgroup

\vfill\pagebreak


\bibliographystyle{IEEEbib}
\bibliography{strings,refs}

\end{document}